\documentclass[%
 aip,
 amsmath,amssymb,
 reprint,%
]{revtex4-1}

\usepackage{graphicx}
\usepackage{dcolumn}
\usepackage{bm}

\usepackage[utf8]{inputenc}
\usepackage[T1]{fontenc}
\usepackage{mathptmx}
\usepackage{etoolbox}

\makeatletter
\def\@email#1#2{%
 \endgroup
 \patchcmd{\titleblock@produce}
  {\frontmatter@RRAPformat}
  {\frontmatter@RRAPformat{\produce@RRAP{*#1\href{mailto:#2}{#2}}}\frontmatter@RRAPformat}
  {}{}
}%
\makeatother
\begin{document}

\preprint{AIP/123-QED}

\title[]{Non-Hermitian control of localization in mosaic photonic lattices}
\author{Stefano Longhi}
 \email{stefano.longhi@polimi.it}
 \altaffiliation[Also at ]{IFISC (UIB-CSIC), Instituto de Fisica Interdisciplinar y Sistemas Complejos - Palma de Mallorca, Spain}
\affiliation{ Dipartimento di Fisica, Politecnico di Milano, Piazza L. da Vinci 32, I-20133 Milano, Italy
}%


\date{\today}

\begin{abstract}
Exploring the deep insights into localization,
disorder, and wave transport in non-Hermitian systems is an emergent area of research of relevance in
different areas of physics. Engineered photonic lattices, with spatial regions of optical gain and loss, provide a prime and simple physical platform for tailoring non-Hermitian Hamiltonians 
and for unveiling the intriguing interplay between disorder and non-Hermiticity. Here it is shown that in mosaic photonic lattices with on-site uncorrelated disorder or quasi-periodic order,
 the addition of uniform loss at alternating sites of the lattice results in the suppression or enhancement of wave spreading, thus providing a simple method for non-Hermitian control of wave transport in disordered systems.
 The results are illustrated by considering discrete-time quantum walks in synthetic photonic lattices.
\end{abstract}

\maketitle

%


Inspired by the concepts of non-Hermitian (NH) physics \cite{rm1,rm2}, in the past two decades NH photonics has emerged as a flourishing area of research,
  enabling 
to mold the flow
of light in unprecedented ways (see, e.g., Refs. \cite{r1,r2,r3,r4,r5,r6,r7,r8,r8bis} and references therein).
In NH systems,
wave transport, localization, and scattering can be deeply
modified as compared to Hermitian systems. 
For example, non-orthogonality of modes and scattering in optical systems with spatial regions of optical gain and loss is responsible for a wide variety of
intriguing effects, such as the appearance of exceptional points, unidirectional scattering, chirality, invisibility, enhanced sensitivity to perturbations, etc. \cite{r1,r2,r3,r4,r5,r6}
An important class of NH systems that is attracting a great interest since long time is provided by systems with disorder or quasiperiodic order, in which the non-Hermitian nature of the Hamiltonian 
can deeply modify the localization and transport properties of waves (see e.g. \cite{r11,r12,r13,r14,r15,r16,r17,r18,r19,r20,r21,r22,r23,r24,r25,r26,r27,r28,r29,r30,r31,r32,r33,r34} and reference therein). 
Recently, the interplay among non-Hermiticity and disorder
has seen a renewed interest, also in connection with the non-trivial spectral topology of NH systems  
 \cite{r35,r36,r37}. To this regard, photonics has provided several experimentally accessible platforms with great flexibility to synthesize
NH models with controllable topology and disorder \cite{r37,r38,r39,r40,r41,r42,r42b}.
 A largely open question is whether and how gain and loss can be harnessed to control localization and wave spreading in disordered systems.
  While application of imaginary gauge fields, i.e. non-reciprocal couplings, is known to prevent Anderson localization and to induce a NH localization-delocalization transition with robust directional transport \cite{r13,r14,r15,r18,r19}, it is not clear whether application of local gain or loss in the system can be harnessed to change the wave transport features in a controllable and simple way. Since in an Hermitian system transport is enabled by extended or weakly localized states of the Hamiltonian, a simple strategy to control wave localization in an Hermitian disordered system sustaining both localized and extended states would be to selectively introduce loss for either extended or localized states, i.e. to selectively control the lifetime of extended and localized states, so as to suppress or enhance wave spreading. While such a method of mode selection could work for clean systems \cite{r42c,r42d}, unfortunately this strategy seems to be hopeless in disordered systems since the wave function profiles of the Hamiltonian depend on the specific realization of disorder in a complex manner.
  
  In this Letter it is shown that, in  a class of mosaic (binary) disordered lattices \cite{r43,r43b,r43c}, in which uncorrelated disorder or commensurate disorder is impressed at alternating sites of the lattice, application of unbalanced losses at alternating sites enables to strategically enhance or suppress wave spreading in the system, thus providing a NH route of  wave localization control without resorting to non-reciprocal couplings. The results are illustrated by considering discrete-time quantum walks in synthetic photonic lattices.
  
 The Hamiltonian of the dissipative tight-binding lattice reads
 \begin{eqnarray}
  H & = & J \sum_n (|n+1 \rangle \langle n|+ | n \rangle \langle n+1|) + \sum_{n} (V_n-i \gamma_n) |n \rangle \langle n|  
 \end{eqnarray}
 where $J$ is the hopping rate between adjacent sites and $V_n$, $\gamma_n$  describe the on-site potential disorder and loss rates, respectively. After letting $|\psi \rangle= \sum_n \psi_n | n \rangle$ for the wave function, the energy spectrum of $H$ is obtained from the discrete Schr\"odinger equation
 \[
 E \psi_n=J(\psi_{n+1}+\psi_{n-1})+(V_n-i \gamma_n) \psi_n
 \]
 For a mosaic (binary) lattice, disorder is applied to odd sites of the lattice solely (sublattice B), i.e. $V_n=0$ for $n$ even. We also assume that the loss rates $\gamma_n$ take only the two values $\gamma_n=\gamma_A$ for $n$ even (sublattice A) and $\gamma_n=\gamma_B$ for $n$ odd (sublattice B); see Fig.1(a) for a schematic.  Given the binary nature of the lattice, it is worth writing the wave function as $| \psi \rangle = \sum_{n} ( a_n |2n \rangle+b_n| 2 n+1 \rangle) $, so that the energy spectrum of the Hamiltonian ${H}$ is defined by the set of coupled equations
 \begin{eqnarray}
 Ea_n & = & J(b_n+b_{n-1})-i \gamma_A a_ n \\
 Eb_n & = &  J(a_n+a_{n+1})+(V_{2n+1}-i \gamma_B) b_ n. 
 \end{eqnarray}
\begin{figure}
\includegraphics[width=8 cm]{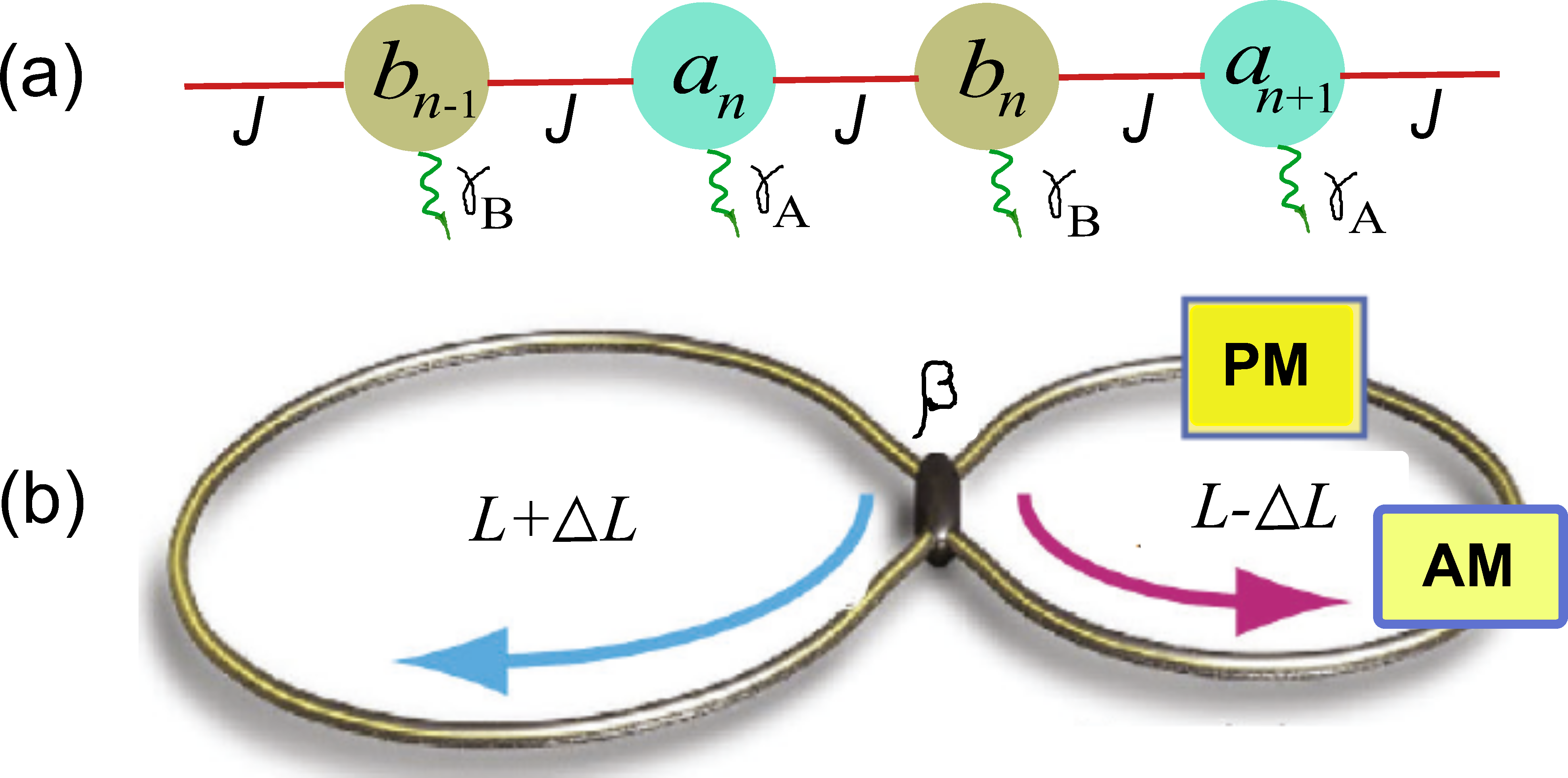}
\caption{(a) Schematic of a mosaic (binary) dissipative lattice with loss rates $\gamma_A$ and $\gamma_B$ in the two sublattices A and B. Disorder is applied at sites of sublattice B solely. $J$ is the coupling rate between adjacent sites.  (b) Schematic of the coupled fiber  loops of slightly unbalance lengths $L \pm \Delta L$ that realize the dissipative mosaic lattice in synthetic dimension. $\beta$ is the coupling angle between the two fiber loops, AM and PM are amplitude and phase modulators.}
\end{figure}
 An interesting result that readily follows from Eqs.(2) and (3) is that, regardless of the specific form of the disordered potential $V_n$,  the energy $E=-i \gamma_A$ always belongs to the spectrum of $H$ and the corresponding wave function is an extended state occupying only the sublattice A, given by $a_n=(-1)^n$ and $b_n=0$. For $E \neq -i \gamma_A$, we can eliminate the amplitudes $a_n$ from Eqs.(2) and (3), 
 \begin{equation}
 a_n=\frac{J(b_n+b_{n-1})}{E+i \gamma_A}
 \end{equation}
 yielding the following recursive equation for the amplitudes $b_n$ in sublattice B
 \begin{equation}
 J^2(b_{n+1}+b_{n-1})+ W_n b_n =[(E+i \gamma_B)(E+i \gamma_A)-2J^2 ] b_n
 \end{equation}
 where we have set
 \begin{equation}
W_n \equiv (E+i \gamma_A) V_{2n+1}.
\end{equation}
Equation (5) can be regarded as the spectral problem of a tight-binding lattice with hopping amplitude $J^2$ and with an effective {\em energy-dependent} on-site disordered potential $W_n$. Interestingly, for an energy $E$ close to $-i \gamma_A$, 
the effective potential is weak (vanishing as $E+i \gamma_A$), so that the corresponding wave functions are expected to be either extended states or weakly-localized states (depending on the nature of disorder $V_n$).  Hence, the lifetime of such extended (or weakly localized) eigenstates of ${H}$ -- given by the inverse of the imaginary part (in modulus) of the energy -- is $ \simeq 1/\gamma_A$.  Also, from Eq.(4) it follows that for such wave functions the excitation is mostly localized in sublattice A. Conversely, for energies such that $E+i \gamma_A$ is far from zero, the effective potential $W_n$ is not anymore weak and the corresponding wave functions are expected to be 
moderately or strongly localized (for reasonable strong potential disorder $V_n$), occupying both sublattices A and B rather generally. Hence, the localized eigenstates of ${H}$ should have a lifetime that lies between $ 1/ \gamma_A$ and $ 1/ \gamma_B$. This means that, for $\gamma_A < \gamma_B$ ($\gamma_A> \gamma_B$) the extended (or weakly localized) states have a longer (shorter) lifetime than the localized states, resulting in an enhancement (suppression) of wave spreading for a rather arbitrary initial localized excitation of the system. Therefore, regardless of the specific form and realization of disorder, selective application of losses at either sublattice A or B should provide a simple and viable route to wave spreading control, i.e. to either suppress of enhance transport in the lattice. We mention that, as compared to the NH delocalization transition obtained by applying imaginary gauge fields \cite{r3,r13,r14}, our method is much simpler since it does not require to make mode coupling asymmetric. To get deeper insights into such a strategical control of wave spreading, let us consider two paradigmatic models of disorder: the incommensurate disorder and the uncorrelated disorder. 

The first case corresponds to the quasi-periodic potential $V_n=2 V_0 \cos (2 \pi \alpha n+ \theta)$, where $\alpha$ is irrational Diophantine and $\theta$ an arbitrary phase. In the Hermitian limit $\gamma_A=\gamma_B=0$, this model exhibits exact mobility edges near $E=0$, separating extended and localized wave functions \cite{r42}. In fact, in this case one has $W_n=2 EV_0 \cos (4\pi \alpha n +2 \pi \alpha+ \theta)$ and Eq.(5) describes the famous Aubry-Andr\'e model \cite{r44}, which is known to have all extended states for $|EV_0|<J^2$ and all localized states for $|EV_0|>J^2$. Further, the energy-dependent Lyapunov exponent $\gamma(E)$, i.e. inverse of localization length of the eigenstates $b_n$ of Eq.(5), reads
\begin{equation}
\gamma(E)=
\left\{
\begin{array}{cc}
0 & |E|<J^2 /V_0 \\
{\rm ln}(|E|V_0/J^2) & |E|>J^2 /V_0
\end{array}
\right.
\end{equation}
Therefore a narrow energy interval, centered at $E=0$ and of width $2 J^2/V_0$, corresponds to extended states that permit ballistic wave spreading in the lattice. When we introduce losses in the system, we expect that wave spreading is suppressed for $\gamma_A> \gamma_B$, while it is enhanced when $\gamma_A<\gamma_B$. The lifetime of the eigenfunctions of ${H}$ is the inverse of the imaginary part (in modulus) of the energy $E$, whereas their degree of localization  is measured by the inverse participation ratio (IPR), which for a normalized wave function on a lattice of size $L$ is defined as $IPR=\sum_n(|a_n|^4+|b_n|^4)$. Note that $0<IPR \leq 1$; for an extended or weakly-localized wave functions, the IPR takes a small value (of order $\sim 1/L$), whereas for a tightly confined wave function the IPR takes a finite value, with $IPR=1$ when excitation is localized in a single site. 
 An example of enhancement and suppression of wave spreading  for incommensurate disorder is shown in Fig.2. We assumed a lattice of size $L=500$ with open boundary conditions  and parameter values $J=1$, $V_0=2.5$, $\theta=0$ and $\alpha= (\sqrt{5}-1)/2$.  The figures depict the numerically-computed energy spectrum of ${H}$ in complex energy plane [panels (a1), (b1) and (c1)], IPR of corresponding eigenstates [panels (a2), (b2) and (c2)], and temporal evolution of the wave amplitudes $|\psi_n(t)|$ corresponding to single-site excitation $\psi_n(0)=\delta_{n,0}$ [panels (a3), (b3) and (c3)]. Panels (a) refer to the Hermitian regime $\gamma_A=\gamma_B=0$; in panels (b)  $\gamma_A=0.05$, $\gamma_B=0$, corresponding to suppression of wave spreading (dynamic localization); in panels (c) $\gamma_A=0$, $\gamma_B=0.05$, corresponding to wave spreading enhancement. Note that, according to the theoretical analysis, by flipping the loss rates in the two sublattices the lifetimes of extended states, near the energy ${\rm {Re}}(E)=0$, change from low [panel (b1)] to high values [panel (c1)] as compared to the lifetimes of localized states, thus explaining wave spreading suppression and enhancement. The wave spreading in the lattice can be experimentally detected by measuring the temporal evolution of the second moment $\sigma^2(t)=(\sum_n n^2 | \psi_n(t)|^2) / ( \sum_n | \psi_n(t)|^2)$ for an initial excitation of site $n=0$ (see for instance \cite{r31}). Figure 2(d) shows the temporal evolution of $\sigma^2(t)$ on a log scale in the three different regimes, clearly showing the suppression (curve 1) or enhancement (curve 2) of wave spreading as compared to the Hermitian case (curve 3).

\begin{figure}
\includegraphics[width=8.5 cm]{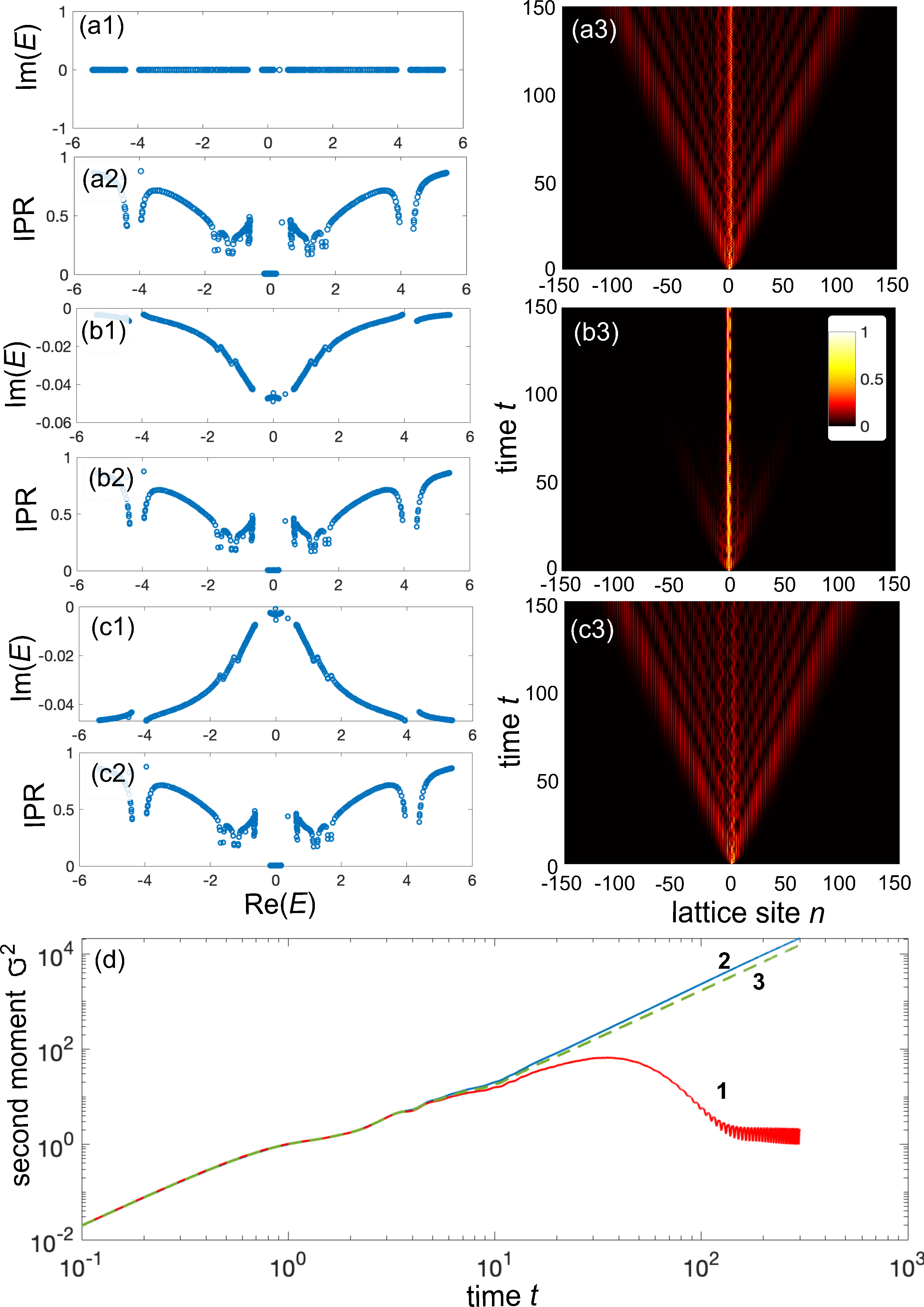}
\caption{Non-Hermitian control of wave spreading in a mosaic dissipative lattice with incommensurate potential. In (a) $\gamma_A=\gamma_B=0$, corresponding to the Hermitian regime. In (b) $\gamma_A=0.05$, $\gamma_B=0$, corresponding to suppression of wave spreading (dynamic localization);  in (c) $\gamma_A=0$, $\gamma_B=0.05$, corresponding to wave spreading enhancement. Other parameter values are given in the text. The energy spectra of the Hamiltonian in complex energy plane are shown in panels (a1), (b1) and (c1); the IPR of corresponding eigenfunctions are depicted in panels (a2), (b2) and (c2). Panels (a3), (b3) and (c3) show on a pseudocolor map wave spreading dynamics in the lattice for initial excitation of site $n=0$. Panel (d) shows the corresponding temporal evolution of the second moment $\sigma^2(t)$ on a log scale (curve 1: $\gamma_A=0.05$, $\gamma_B=0$; curve 2: $\gamma_A=0$, $\gamma_B=0.05$; curve 3: $\gamma_A=\gamma_B=0$).}
\end{figure}

The second case corresponds to uncorrelated  disorder, i.e. $V_n$ are independent random variables with the same probability density  $p(V)$. In the Hermitian limit $\gamma_A=\gamma_B=0$, Eq.(5) describes the usual problem of 
Anderson localization in a one-dimensional lattice with random disorder $W_n$ and thus, for any energy $E \neq 0$, all eigenstates $b_n$ are Anderson localized, regardless of how weak is the effective random potential $W_n$ \cite{And1,And2,And2b} . This means that, contrary to the incommensurate case discussed above and displaying mobility edges, the Lyapunov exponent $\gamma(E)$ vanishes only at $E=0$, and $\gamma(E)>0$ for $E \neq 0$. This situation is similar to the famous random dimer model \cite{dimer}, indicating that a set of weakly-localized states, with a diverging localization length and density of states, accumulates toward $E=0$.
The specific form of $\gamma(E)$ depends on the probability density $p(V)$ of disorder and can be given analytically in very special cases. For example, for the Cauchy distribution, $p(V)=(\delta / \pi)(\delta^2+V^2)^{-1}$, the statistically-averaged Lyapunov exponent reads \cite{And2b}
\begin{equation}
\gamma(E)={\rm cosh}^{-1}  \left( \frac{\sqrt{(E^2-4J^2)^2+\delta^2 E^2} +\sqrt{E^4+\delta^2 E^2}}{4J^2} \right)
\end{equation}
vanishing like $\gamma(E) \sim |E|^{1/2}$ as $E \rightarrow 0$,
whereas for any distribution with a finite variance $ \langle \epsilon^2 \rangle$, such as the uniform distribution, the behavior of $\gamma(E)$ can be calculated analytically in the neighbor of $E=0$ using a perturbative method \cite{And3}, and reads
\begin{equation}
\gamma(E) \simeq 0.2893 \frac{ \langle \epsilon^2 \rangle^{1/3} |E|^{2/3} }{J^{4/3}}
\end{equation}
vanishing like $\gamma(E) \sim |E|^{2/3}$ as $E \rightarrow 0$.
We note that, even though the Hamiltonian ${H}$ for uncorrelated disorder has an almost pure point spectrum, with the exception of the extended state at $E=0$, like in the random dimer model \cite{dimer} wave spreading and sub-ballistic transport in the lattice is 
allowed by the set of weakly localized states with diverging localization length near $E=0$ (see also \cite{Longhi} for the case of non-random potentials). An example of wave spreading suppression and enhancement for a uniform probability distribution, $p(V)=1/ \delta$ for $|V|< \delta /2$ and $p(V)=0$ for $|V|> \delta /2$, obtained by adding losses in either sublattices A or B, is shown in Fig.3. Note that, as compared to the incommensurate disorder of Fig.2, for random disorder the wave spreading enhancement is larger [compare curves 2 and 3 in Figs.2(d) and 3(d)]. This behavior can be explained as follows. In the incommensurate case and in the Hermitian limit, the Hamiltonian $H$ shows mobility edges with extended states enabling ballistic transport, and the addition of losses at odd lattice sites, yielding a decrease of the lifetime of localized states, does not substantially increase the wave spreading of the system. On the other hand, in the random potential case in the Hermitian limit the spectrum is almost pure point and wave spreading is sub-ballistic. In this case wave spreading greatly benefits from the decrease of the lifetime of strongly-localized states, as compared to weakly-localized states, by the addition of losses at odd sites of the lattice.

\begin{figure}
\includegraphics[width=8.5 cm]{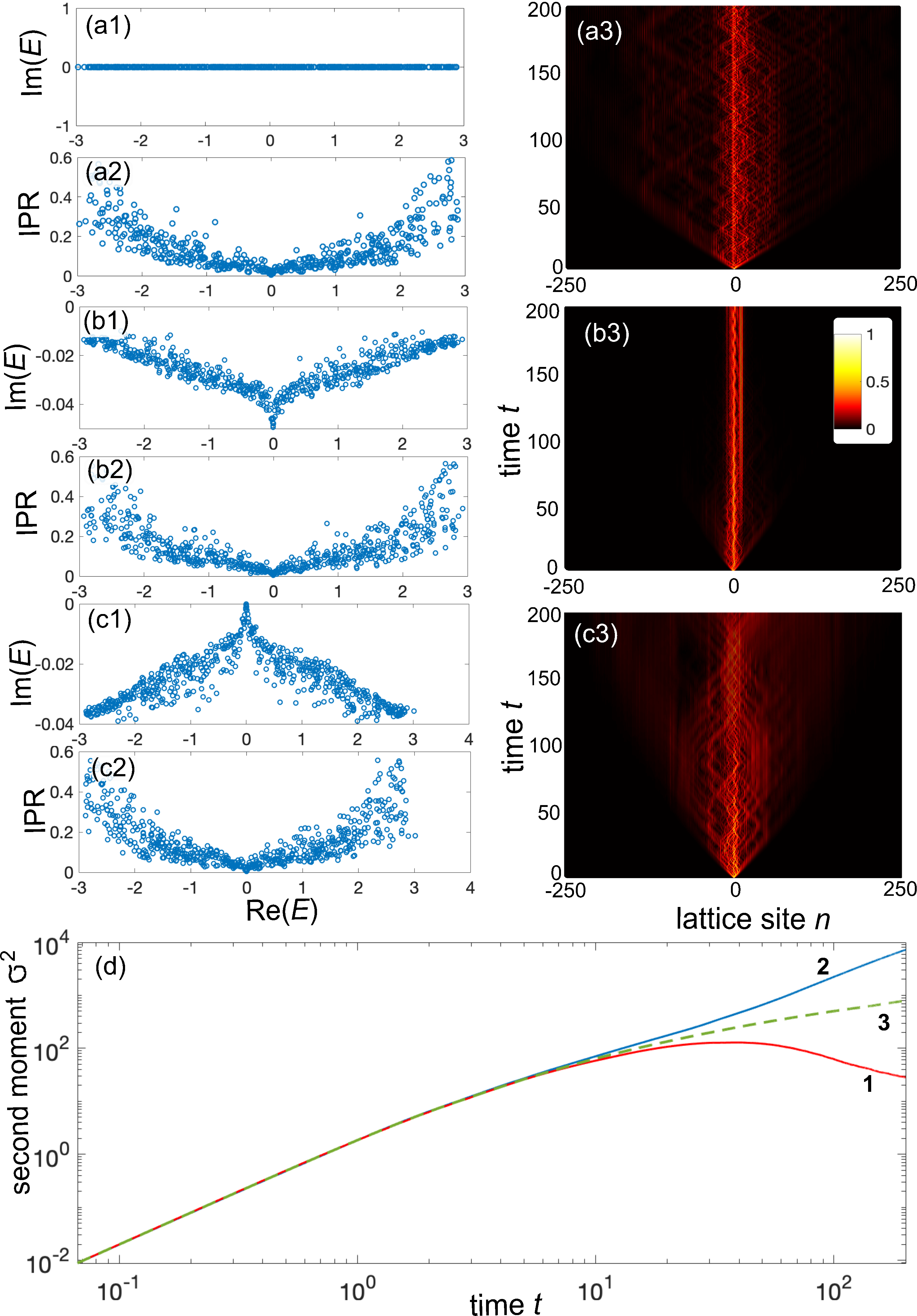}
\caption{Same as Fig.2 but for a random didorder with uniform probability density $p(V)$. Parameter values are $J=1$, $\delta=4$ with $\gamma_{A}=\gamma_B=0$ in (a) (Hermitian regime), $\gamma_{A}=0.05$, $\gamma_B=0$ in (b), and  $\gamma_{A}=0$, $\gamma_B=0.05$ in (c). Panels (a), (b) and (c) depict energy spectrum, IPR and wave spreading dynamics for a single realization of disorder, whereas the curves in (d) are obtained after statistical average over 200 realizations of disorder.}
\end{figure}

An experimentally feasible platform to realize NH photonic lattices with controllable disorder in synthetic space is provided by discrete-time quantum walks of optical pulses in coupled fiber loops
 (see e.g. \cite{r31,r37,r40,r41,fiber1,fiber2,fiber3,fiber4} and references therein). The system consists of two fiber loops of slightly different lengths $(L \pm \Delta L)$ that are connected by a fiber coupler with a coupling
angle $\beta$, as schematically shown in Fig.1(b). A phase and amplitude modulators are
placed in one of the two loops, which provide a desired control of the phase and amplitude of the traveling pulses.
Light evolution is described by the set of discrete-time coupled-mode equations \cite{r31,r37,Longhi,fiber1,fiber2,fiber3}
 \begin{eqnarray}
 u^{(m+1)}_n & = & \left(   \cos \beta u^{(m)}_{n+1}+i \sin \beta v^{(m)}_{n+1}  \right)  \exp (-2i V_n-2\gamma_n) \; \;\; \\
 v^{(m+1)}_n & = & \left(   \cos \beta v^{(m)}_{n-1}+i \sin \beta u^{(m)}_{n-1}  \right)
 \end{eqnarray}
  where $u_n^{(m)}$ and $v_n^{(m)}$ are the pulse amplitudes at discrete time step $m$ and lattice site $n$ in the two fiber loops, and $ V_n$, $ \gamma_n$  are the phase and amplitude terms impressed by the phase and amplitude modulators, respectively. Assuming a coupling angle $ \beta$ close to $\pi /2$ and for weak phase and amplitude modulations, the light dynamics can be effectively described by a continuous-time model [Eq.(1)], with the discrete time $m$ replaced by a continuous time variable $t$ and with a hopping amplitude $J=  \pm (1/2) \cos\beta$ \cite{fiber4,Longhi,Longhi2}. Therefore, in such a limit the discrete-time quantum walk [Eqs.(10) and (11)] realizes the Hamiltonian (1) of the mosaic lattice with controllable disorder $V_n$ and alternating loss rates $\gamma_n$, which are set by the phase and amplitude modulators.
  An example of wave spreading suppression and enhancement in the discrete-time quantum walk for an incommensurate potential is shown in Fig.4. The figure depicts the numerically-computed evolution of the normalized occupation probabilities
  $P_n^{(m)}=|u_n^{(m)}|^2+|v_n^{(m)}|^2 / \sum_n (|u_n^{(m)}|^2+|v_n^{(m)}|^2) $ 
  at successive discrete time steps $m$ [panels (a), (b) and (c)], and corresponding second moment $\sigma^2(m)=\sum_n n^2 P_n^{(m)}$ [panel (c)] for the incommensurate potential $V_n=V_0 \cos ( 2 \pi \alpha n)$; parameter values are $\beta =0.98 \times \pi/2$, $V_0=0.02$, and $ \alpha=(\sqrt{5}-1)/2$. Single pulse excitation at site $n=0$ is assumed, corresponding to $u_n^{(0)}=\delta_{n,0}$ and $v_n^{(0)}=0$. In Fig.4(a) the system is Hermitian ($\gamma_A=\gamma_B=0$), in Fig.4(b) $\gamma_A=0.02$ and $\gamma_B=0$, corresponding to suppression of wave spreading, in Fig.4(c) $\gamma_A=0$ and $\gamma_B=0.02$, corresponding to enhancement of wave spreading. Similar results are obtained by assuming random (rather than incommensurate) disorder for the phase $V_n$. 
 
\begin{figure}
\includegraphics[width=8.5 cm]{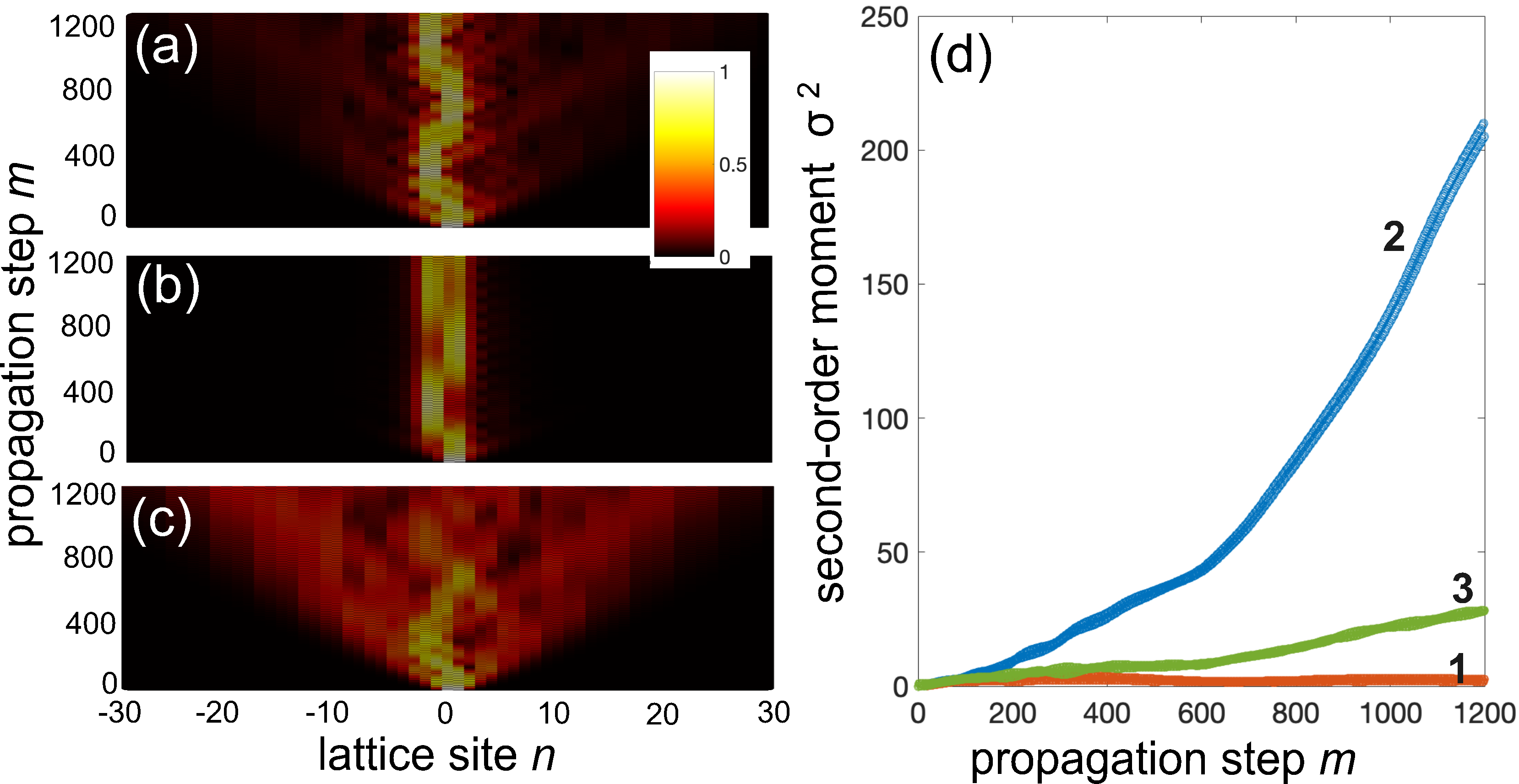}
\caption{Non-Hermitian control of light spreading in a synthetic photonic lattice realized by the coupled fiber loop setup of Fig.1(b). Coupling angle $\beta=0.98 \times \pi/2$, phase modulation (incommensurate potential) $V_n=2 V_0 \cos (2 \pi \alpha n)$ at odd sites with $V_0=0.02$, $\alpha=(\sqrt{5}-1)/2$, and loss modulation $\gamma_n=\gamma_A$ at even sites, $\gamma_n=\gamma_B$ at odd sites. Panels (a-c) show the numerically-computed evolution of the occupation probability $P_n^{(m)}$ versus discrete time step $m$ on a pseudo color map for initial single-pulse excitation of the loops. In (a) $\gamma_A=\gamma_B=0$ (Hermitian lattice), in (b) $\gamma_A=0.02$ and $\gamma_B=0$, corresponding to suppression of wave spreading (dynamical localization), in (c)  $\gamma_A=0$ and $\gamma_B=0.02$, corresponding to wave spreading enhancement. The behavior of the second moment $\sigma^2$ versus $m$ in the three cases in shown in panel (d). Curve 1: $\gamma_A=0.02$, $\gamma_B=0$; curve 2: $\gamma_A=0$, $\gamma_B=0.02$; curve 3: $\gamma_A=\gamma_B=0$.}
\end{figure}

In conclusion, a simple and feasible method of NH wave spreading control in a class of disordered mosaic lattices, based on application of alternating local losses in the lattice, has been theoretically proposed and demonstrated in numerical simulations for different types of disorder. The technique enables to strategically enhance or suppress wave spreading in the lattice in a simple and universal way, thus providing a NH route of  wave localization control without resorting to non-reciprocal couplings. The results have been illustrated by considering discrete-time quantum walks in synthetic photonic lattices, which should provide a feasible photonic platform for the observation of loss-induced control of localization.


\begin{acknowledgments}
The author acknowledges the Spanish State Research Agency, through 430
the Severo Ochoa and Maria de Maeztu Program for Centers 431
and Units of Excellence in R\&D (Grant No. MDM-2017- 432
0711).
\end{acknowledgments}

\section*{Data Availability Statement}
Data available on request from the author.




\end{document}